\documentstyle[pra,aps,twocolumn,epsf]{revtex}

\begin{document}
\title{Experimental study of the momentum correlation of a pseudo-thermal field in the photon counting regime}
\author{Giuliano Scarcelli, Alejandra Valencia, and Yanhua
Shih}
\address{Department of Physics, University of Maryland, Baltimore
County, Baltimore, Maryland 21250} \maketitle\date{17 March, 2003}

\begin{abstract}
Thermal (or pseudo-thermal) radiation has been recently proposed
for imaging and interference types of experiments to simulate
entangled states.  We report an experimental study on the momentum
correlation properties of a pseudo-thermal field in the photon
counting regime. The characterization and understanding of such a
light source in the context of two photon physics, especially its
similarities and differences compared to entangled two-photon
states, is useful in gaining knowledge of entanglement and may
help us to assess the real potential of applications of chaotic
light in this context.
\end{abstract}

\vspace{1cm}The study of thermal radiation has been historically
important in optics \cite{mandel}. Before the invention of the
laser, thermal sources were probably the only light sources
available in the laboratory for optical experiments and
observations.  Although coherent light has taken the major role in
modern optics laboratory, thermal, or pseudo-thermal, sources are
still important for the understanding of fundamental concepts,
such as optical coherence and photon statistics
\cite{mandel,hanbury,fano,glauber1}.

Recently it has been theoretically proposed to use thermal, or
quasi-thermal, radiation \cite{thermaltheory1} in the context of
two-photon imaging and interference \cite{pittman}. The proposal
fits in the lively ongoing discussion about the possibility of
using classically correlated light sources to simulate the
behavior of entangled photon states. In the past two years, in
fact, successful attempts to simulate some of the surprising
features of entangled states with classically correlated light
have been made \cite{bennink}, leading to an interesting
discussion about which features are peculiar to entanglement, and
to what extent a classical correlation can simulate
them\cite{discussion}. Furthermore it has been argued that thermal
light exhibits exactly the kind of correlations necessary to
simulate, at least qualitatively, most of the results obtained in
the past through entangled photon states \cite{thermaltheory1}.

In this paper we report an experimental study on the momentum
correlation properties of thermal radiation in the photon counting
regime. A theoretical analysis is also proposed to clarify the
physics behind the observed phenomenon, especially the
similarities and differences with entangled two-photon states. It
may not be quite correct to use the terminology ``correlation" to
describe the physics behind the experimental observation. In fact,
the observed result comes naturally from the \textit{complete
absence of correlation}, as is hinted by the same name of chaotic
light.

In principle the term ``thermal radiation" should refer only to
radiation coming from a blackbody in thermal equilibrium at some
temperature $T$. But in reality, some characteristics of true
thermal fields, like the extreme shortness of their coherence
time, have imposed serious obstacles to their use in actual
experiments, and therefore since the early days of quantum optics
there has been a great interest in the realization of more
utilizable sources that could simulate the behavior of true
thermal fields (gas discharge lamps, randomized lasers etc.). We
usually describe these kind of sources as pseudo-thermal; they are
actually all chaotic light sources that can be modelled as a
collection of independent atoms emitting radiation randomly
\cite{loudon}. The most commonly used among the pseudo-thermal
fields is the one developed by Martienssen and Spiller
\cite{martienssen} in the mid of $1960$ 's. The principle of the
generation is very simple: coherent laser radiation is focused on
a rotating ground glass disk so the scattered radiation is chaotic
with a Gaussian spectrum.

We start by describing the light source used in this experiment.
We experimentally studied the characteristics of the source to be
sure that it indeed behaved like a thermal field. To this purpose
we used an experimental setup similar to the one that was later
used to perform the momentum correlation measurement
(Fig.~\ref{setup}). First we focused on the temporal behavior and
repeated a historical experiment of Arecchi et al \cite{arecchi}.
The point of the experiment was to study the time distribution of
the coincidence counts coming from the source under study.
Fig.~\ref{mca} shows the results of the measurements. Basically we
plot a histogram of the number of coincidence counts versus the
time difference of the photon registration clicks from the two
detectors. The temporal behavior of our source is similar to the
one obtained with light coming from stars and gas discharge lamps
\cite{hanburybook,scarl,phillips}, but with a higher coherence
time (around $1 \mu s$, mainly determined by the rotation speed of
the disk and the properties of the grit of the ground glass). In
practice this graph is also useful for ``calibrating" the
coincidence time window; in fact in the rest of the experiment we
set a window of about $600$ ns around the peak of Fig.~\ref{mca}
and measured the number of coincidence counts in that window. Then
we studied the spatial behavior of the coincidence counts using
the setup in Fig.~\ref{setup}. The point of this measurement is to
study the behavior of the spatial correlations when the size of
the source is varied. This kind of measurement is basically the
one used to determine the angular size of the stars in
Hanbury-Brown and Twiss stellar interferometry. Experimentally we
fixed detector $D_{1}$ in one position and scanned horizontally
the position of detector $D_{2}$ for different sizes of the source
obtained by focusing or defocusing the He-Ne laser onto the
rotating disk. The results are shown in Fig.~\ref{spotsize}. As
expected, the bigger is the angular size of the source, the
smaller is the width of the correlation function, in agreement
with stellar interferometric predictions.

These two measurements confirmed that our source simulated both
spatially and temporally the behavior of thermal radiation. We
then proceeded on measuring the momentum-momentum correlation of
the field. The setup for the measurement is shown in
Fig.~\ref{setup}. After the source, a lens ($L$) of focal length
$f=75 cm$ was inserted, and then the radiation was split by a
non-polarizing cube beam splitter ($l=25 mm$). In the focal plane
of the lens $L$ we put the detection system composed of two
horizontally displaceable fibers ($60 \mu m$ diameter) connected
to two single-photon counting modules. The output pulses from the
two detectors were then sent to an electronic coincidence circuit.
The setup is designed to study the momentum properties of the
fields, in fact the lens $L$ was close to the source and the
detectors were placed in its focal plane to ensure the far-field
condition. In this situation, in principle, the fiber tip
(detector) scans the transverse wavevector plane, or transverse
momentum plane, of the field. Therefore by scanning the fibers in
the horizontal direction continuous information about the behavior
of each k-vector (momentum) of the field will be obtained.

We performed three sets of measurements for three different
position of detector $D_{1}$; in every measurement we kept the
position of $D_{1}$ fixed and scanned horizontally the position of
detector $D_{2}$.  The results are shown in Fig.~\ref{coin}. It is
clear that the correlation function ``followed" the movement of
detector $D_{1}$; a shift in the position of $D_{1}$, causes a
shift in the same direction of the correlation function. The
natural deduction from this measurement is that the field is
indeed correlated in momentum: in fact every time  a position of
$D_{1}$ is fixed, one k-vector from the source is selected,
therefore the behavior of the coincidence counts show that only
one k-vector arriving at $D_{2}$ is correlated to the selected
k-vector arriving at $D_{1}$. In the central part of
Fig.~\ref{coin} the single counts are reported, notice that (1)
the single counts are flat over the entire range of detection
suggesting the completely random transverse distribution of the
field; (2) the low level of photon counts ($60000$ per second)
suggests that we were effectively operating in the condition of
only two photons present in the time duration of interest.

The phenomenological conclusion of these measurements would then
be that the pseudo-thermal field we have measured is indeed
correlated in momentum. This is a property that makes this source
particularly interesting to study in comparison with entangled
photon pairs produced in Spontaneous Parametric Down Conversion
(SPDC) in the context of two-photon imaging and interference
features \cite{pittman,thermaltheory1}. However, the predicted,
and now experimentally measured, analogy between the effects
obtained with entangled photons and thermal radiation has to be
analyzed very carefully. In order to clarify the point let us
justify theoretically the measured experimental effect. The
quantity that rules the experiments involving coincidence
measurements is the second-order Glauber correlation function
\cite{glauber}:

\begin{eqnarray}\label{G2}
G^{(2)}(t_{1},\textbf{x}_{1}; t_{2},\textbf{x}_{2})  \equiv \hspace{5cm} \\
\nonumber \equiv Tr[\hat{\rho} E_{1}^{(-)}(t_{1},
\textbf{x}_{1})E_{2}^{(-)}(t_{2}, \textbf{x}_{2})
\hspace{2cm}\\\nonumber E_{2}^{(+)}(t_{2},
\textbf{x}_{2})E_{1}^{(+)}(t_{1}, \textbf{x}_{1})].
\end{eqnarray}
Where $\textbf{x}_{i}$ is a vector in the plane of the $i^{th}$
detector perpendicular to the direction of propagation of the
radiation, $\hat{\rho}$ represents the density matrix of the
quantum state under consideration, and $E_{i}^{(\pm )}(t_{i},
\textbf{x}_{i})$, $i=1,2$, are positive-frequency and
negative-frequency components of the field at detectors $D_{1}$
and $D_{2}$ that can be written as:
\begin{eqnarray}\label{field}
E^{(+)}_{i}(t_{i},\textbf{x}_{i})= \sum_{\textbf{k}}
\sqrt{\frac{\hbar \omega}{2 \epsilon_{0} V}}
g_{i}(\textbf{x}_{i},\textbf{k}) a_{_{\textbf{k}}} e^{-i\omega
t_{i}}.
\end{eqnarray}
Here $g_{i}(\textbf{x}_{i},\textbf{k})$ stands for the Green's
function associated with the propagation of the field towards the
$i^{th}$ detector\cite{rubin}. Plugging the expression of the
electric fields in Eq.~\ref{G2}, we obtain the equal time
second-order Glauber correlation function in the following form:
\begin{eqnarray}\label{G2sums}
\nonumber G^{(2)}(\textbf{x}_{1};\textbf{x}_{2})  \propto \hspace{10cm}\\ \vspace{1cm} \sum_{\textbf{k}}\sum_{\textbf{k}'}\sum_{\textbf{k}''}\sum_{\textbf{k}'''}g_{1}^{*}(\textbf{x}_{1},\textbf{k})g_{2}^{*}(\textbf{x}_{2},\textbf{k}')\hspace{7cm} \\
\nonumber \times
g_{2}(\textbf{x}_{2},\textbf{k}'')g_{1}(\textbf{x}_{1},\textbf{k}''')Tr[\hat{\rho}
a^{\dag}_{\textbf{k}}a^{\dag}_{\textbf{k}'}a_{\textbf{k}''}a_{\textbf{k}'''}]\hspace{3.5cm}
\end{eqnarray}
Let us examine the last term of Eq.~\ref{G2sums}, in particular
the density matrix of the state of the radiation we are studying.
Usually the density matrix that is used for single mode thermal
light is expressed in the number state basis as follows
\cite{loudon} \cite{scully}:
\begin{equation}\label{Rhosingle}
\hat{\rho}= \sum_{n}\frac{\langle n \rangle^{n}}{(1+\langle n
\rangle)^{1+n}}|n \rangle\langle n|
\end{equation}
In the multimode case we can take $| \{ n_{\textbf{k}} \} \rangle
$  as basis vectors, where the symbol $\{ n_{\textbf{k}} \}$
indicates a set of number of photons
$n_{\textbf{k}_{1}},n_{\textbf{k}_{2}},n_{\textbf{k}_{3}}$ etc. in
the modes $\textbf{k}_{1},\textbf{k}_{2},\textbf{k}_{3}$
respectively, and  $| \{ n_{\textbf{k}} \} \rangle $ are product
states of the number states $|n_{\textbf{k}}\rangle$ of each mode
$\textbf{k}$. In thermal and chaotic light the different modes of
the field are independent \cite{loudon}, thus the total density
matrix will be the product of the contributions from the different
modes as follows:
\begin{equation}\label{Rhomulti}
\hat{\rho}=\sum_{\{n_{\textbf{k}}\}}|\{n_{\textbf{k}}\}\rangle\langle\{n_{\textbf{k}}\}|
\prod_{\textbf{k}}\frac{\langle
n_{\textbf{k}}\rangle^{n_{\textbf{k}}}}{(1+\langle
n_{\textbf{k}}\rangle)^{1+n_{\textbf{k }}}}\end{equation} where
$\langle n_{\textbf{k}} \rangle$ represent the average number of
photons in mode $\textbf{k}$.

A comparison between the density matrix that describes the thermal
radiation and the state of the entangled photons from SPDC can
readily clarify the intrinsic difference between the two sources.
While in the thermal case it is intuitive and apparent from
Eq.~\ref{Rhomulti} that the different modes of the radiation are
completely independent, it is immediately clear from the state of
the SPDC radiation that there is perfect correlation in momentum
exhibited by the source.

The trace in Eq.~\ref{G2sums} turns out to be:
\begin{equation}\label{trace2}
Tr[\hat{\rho}
a^{\dag}_{\textbf{k}}a^{\dag}_{\textbf{k}'}a_{\textbf{k}''}a_{\textbf{k}'''}]=\langle
n \rangle^{2}
(\delta_{\textbf{k},\textbf{k}''}\delta_{\textbf{k}',\textbf{k}'''}+\delta_{\textbf{k},\textbf{k}'''}\delta_{\textbf{k}',\textbf{k}''})
\end{equation}
where we have assumed the average number of photons in each mode
to be constant (as is confirmed by the single counts behavior in
Fig.~\ref{coin}). Plugging the previous result in Eq.~\ref{G2sums}
we obtain the following expression for the second-order
correlation function:
\begin{eqnarray}\label{G2int1}
G^{(2)}(\textbf{x}_{1}; \textbf{x}_{2})  \propto \hspace{5cm}
 \\ \nonumber \hspace{-0.5cm} \int d^{2}\textbf{q}\mid
g_{1}(\textbf{x}_{1},\textbf{q})\mid^{2}\int d^{2}\textbf{q}\mid g_{2}(\textbf{x}_{2},\textbf{q})\mid ^{2}\\
\nonumber  \hspace{2cm} +  \mid \int d^{2}\textbf{q}
g_{1}^{*}(\textbf{x}_{1},\textbf{q})g_{2}(\textbf{x}_{2},\textbf{q})\mid^{2
}.
\end{eqnarray}
Where $\textbf{q}$ is the transverse component of the k-vector.
Notice that we have converted the summation of Eq.~\ref{field}
into an integral . It is easy to see that $\int
d^{2}\textbf{q}\mid g_{i}(\textbf{x}_{i},\textbf{q})\mid^{2}$ will
actually lead to the number of single counts $n_{i}$ of the
$i^{th}$ detector. The expression can be further simplified:
\begin{eqnarray}\label{G2int2}
G^{(2)}(\textbf{x}_{1}; \textbf{x}_{2})  \propto 1 +\frac{\mid
\int d^{2}\textbf{q}
g_{1}^{*}(\textbf{x}_{1},\textbf{q})g_{2}(\textbf{x}_{2},\textbf{q})\mid^{2
}}{n_{1}n_{2}}
\end{eqnarray}
Inserting the Green's functions associated to the optical setup
into Eq.~\ref{G2int2} we obtain:
\begin{eqnarray}\label{G2int3}
G^{(2)}(\textbf{x}_{1}; \textbf{x}_{2}) & \propto & 1 +
[\frac{J_{1}(\pi a |\textbf{x}_{1}-\textbf{x}_{2}|/\lambda f)}{\pi
a |\textbf{x}_{1}-\textbf{x}_{2}|/2 \lambda f}]^{2}
\end{eqnarray}
where $J_{1}$ is the Bessel function of the first order and $a$ is
the size of the source. This result explains clearly the
experimental observations of both Fig.~\ref{spotsize} and
Fig.~\ref{coin}. In fact, it predicts that the bigger the size of
the source, the smaller is the width of the correlation, and above
all it predicts that shifting the position of detector $D_{1}$
will cause the observed shift of the correlation function. Notice
that, due to the first term in Eq.~\ref{G2int3}, the maximum
achievable visibility in these kind of experiments is $33 \%$.

Although the theoretical calculations are able to predict the
experimental observations, the physical interpretation of the
experimental results and mathematical formalism remains yet to be
fully explained and understood. To clarify this point, let us
focus for a moment on the following quantities:
\begin{eqnarray}\label{firstorderdef}
\langle a^{\dag}_{\textbf{k}} a_{\textbf{k}'} \rangle & \equiv &
Tr[\hat{\rho} a^{\dag}_{\textbf{k}} a_{\textbf{k}'}]\\ \nonumber
\langle a^{\dag}_{\textbf{k}} a_{\textbf{k}'}
a_{\textbf{k}''}a_{\textbf{k}'''}\rangle & \equiv & Tr[\hat{\rho}
a^{\dag}_{\textbf{k}}
a_{\textbf{k}'}a_{\textbf{k}''}a_{\textbf{k}'''}]
\end{eqnarray}
Substituting into this equation the expression for the density
matrix that describes our source we obtain:
\begin{eqnarray}\label{delta}
\langle a^{\dag}_{\textbf{k}} a_{\textbf{k}'} \rangle & = &
\langle n \rangle \delta_{\textbf{k},\textbf{k}'} \end{eqnarray}
\begin{eqnarray}\label{delta2}
\langle
a^{\dag}_{\textbf{k}}a^{\dag}_{\textbf{k}'}a_{\textbf{k}''}a_{\textbf{k}'''}\rangle
& = & \langle n \rangle^{2}
(\delta_{\textbf{k},\textbf{k}''}\delta_{\textbf{k}',\textbf{k}'''}+\delta_{\textbf{k},\textbf{k}'''}\delta_{\textbf{k}',\textbf{k}''})
\end{eqnarray}
Notice that relations of the kind in Eq.~\ref{delta} are generally
used in statistics to express the absolute \textit{statistical
independence} of the variates under study. In this case therefore
Eq.~\ref{delta} can be interpreted as the mathematical expression
of the \textit{lack of correlation} between different modes of the
field \cite{mandel} and the experimental observations would be a
natural manifestation of this property.

As far as Eq.~\ref{delta2} is concerned, in the context of
entangled photon states, a formally similar equation is usually
considered explicative of the perfect momentum correlation between
the two entangled photons due to the phase matching conditions of
the nonlinear process inside the crystal. The interpretation of
this mathematical relation in terms of correlation between
k-vectors could be misleading in the case of chaotic light.  In
fact, it would employ a terminology peculiar to entangled photons
for a process that is intrinsically different.

Besides the physical interpretation, however, the theoretical and
experimental results reveal that the source under study, and thus
thermal radiation in general, is promising for two-photon
interference type experiments, except for the lower visibility
(33\% maximum)\cite{klyshko}. From a fundamental point of view it
can represent a perfect tool for studying the differences and/or
the analogies between quantum entanglement and its classical
simulation.

In conclusion, we have experimentally studied the far-field
momentum ``correlation" of chaotic light fields and we have
provided a theoretical explanation of the results with a physical
interpretation of the phenomenon. The observed correlation is in
fact the experimental evidence of the complete \textit{absence of
correlation} between different modes of the radiation. The
experiment can help us to assess to what extent thermal or
pseudo-thermal fields might be employed to simulate certain
features of entangled photons in the context of two-photon imaging
and interference.

\begin{figure}[hbt]
\caption{Sketch of the experimental setup} \label{setup}
\end{figure}

\begin{figure}[hbt]
\caption{Histogram of number of joint detection counts vs time
difference of the two photo-electron events. The size of each
channel is $0.3$ ns.} \label{mca}
\end{figure}

\begin{figure}[hbt]
\caption{Normalized second order correlation function vs position
of detector $D_{2}$. The various sets of data (points) are for
different sizes of the pseudo-thermal source (circles $\sim 150
\mu m$; squares $\sim 350 \mu m$; triangles $\sim 550 \mu m$;
diamonds $\sim 1100 \mu m$) and agree with the theoretical
expectation (solid lines).} \label{spotsize}
\end{figure}

\begin{figure}[hbt]
\caption{Normalized second order correlation function vs position
of detector $D_{2}$ for three fixed positions of detector $D_{1}$:
(a) $x_{1}\simeq -1.75 mm$; (b) $x_{1}\simeq 0 mm$; (c)
$x_{1}\simeq 1.75 mm$ (source size $\sim 550 \mu m$).  In part (b)
is shown also a plot of single counts vs position of the detector
for both $D_{1}$ (filled squares) and $D_{2}$ (hollow squares).
The level of the counts is around $60000$ per second.}
\label{coin}
\end{figure}

\centerline{\epsfxsize=2in \epsffile{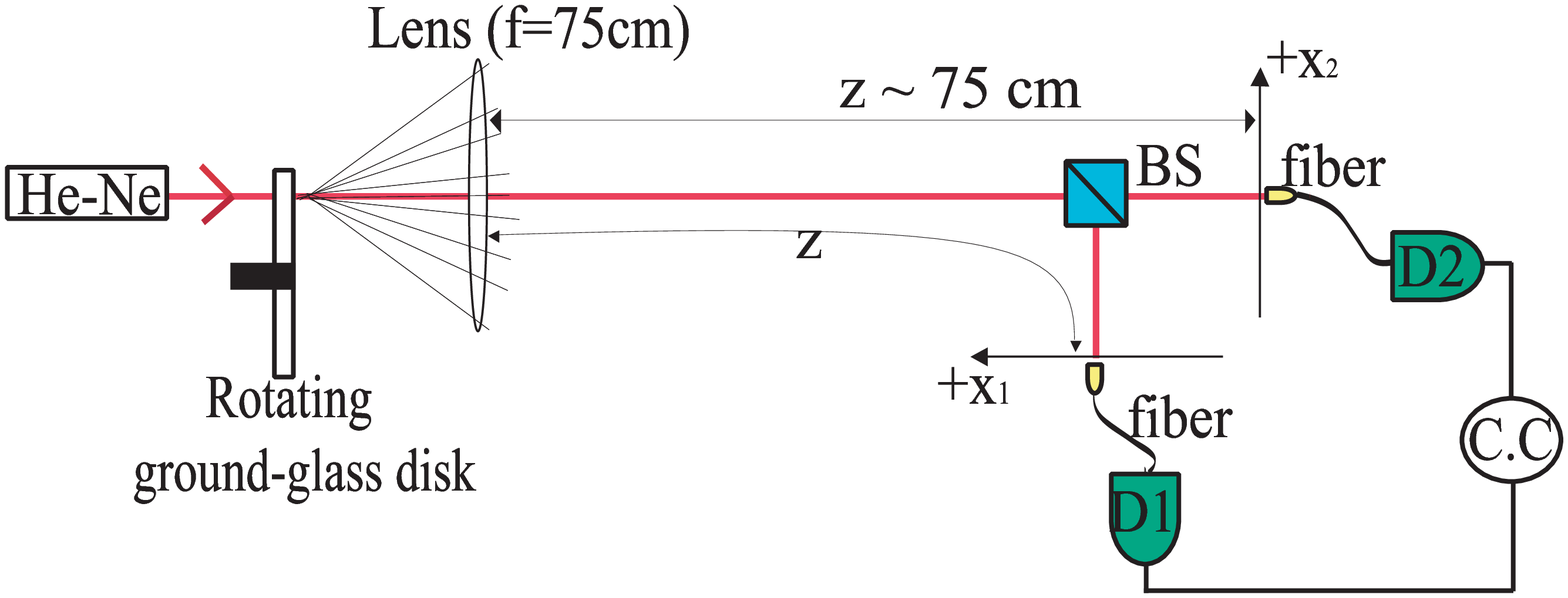}} \vspace{1cm}
Figure \ref{setup}.  Giuliano Scarcelli, Alejandra Valencia, and
Yanhua Shih.

\centerline{\epsfxsize=1.8in \epsffile{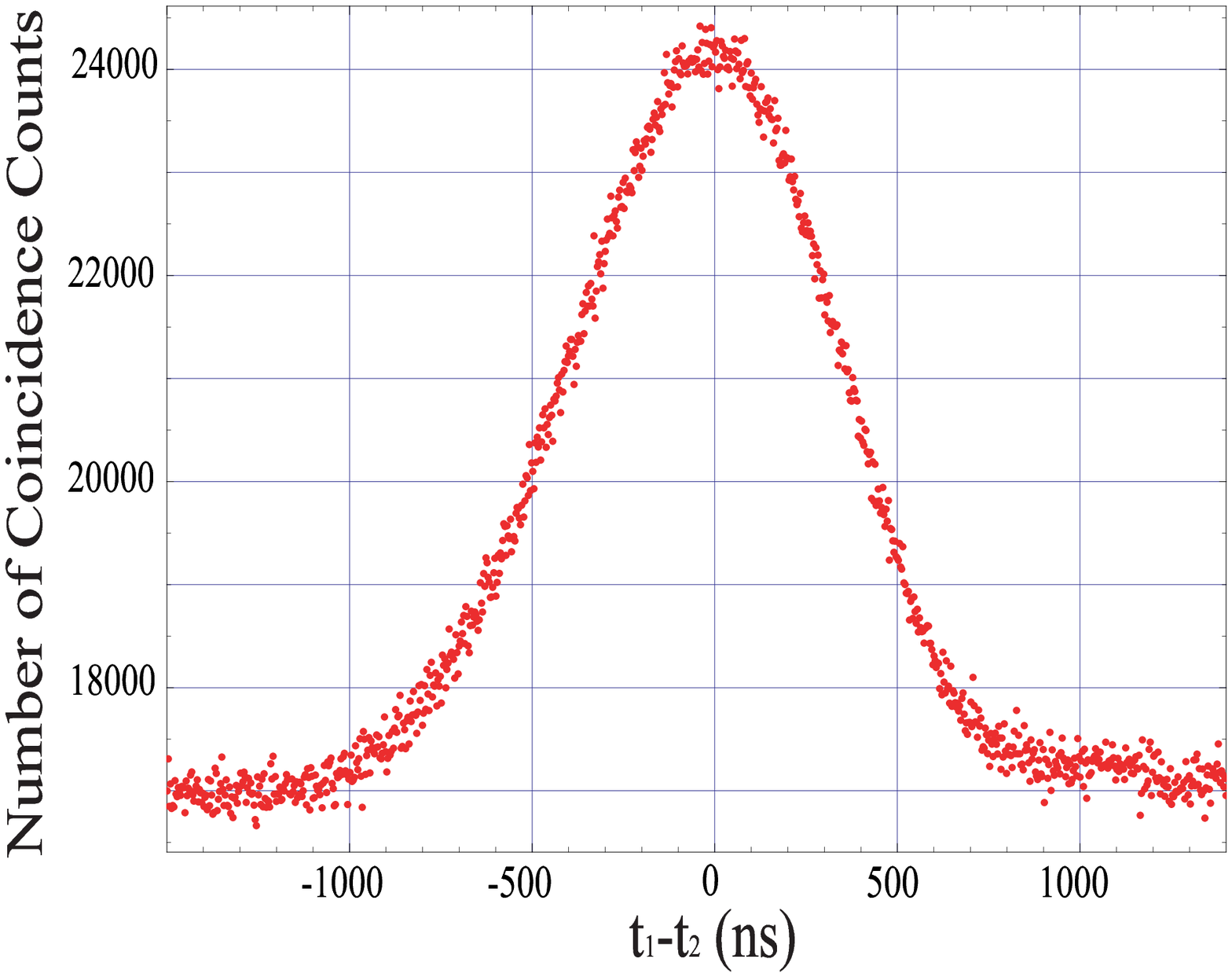}} \vspace{1cm}
Figure \ref{mca}.  Giuliano Scarcelli, Alejandra Valencia, and
Yanhua Shih.

\centerline{\epsfxsize=1.8in \epsffile{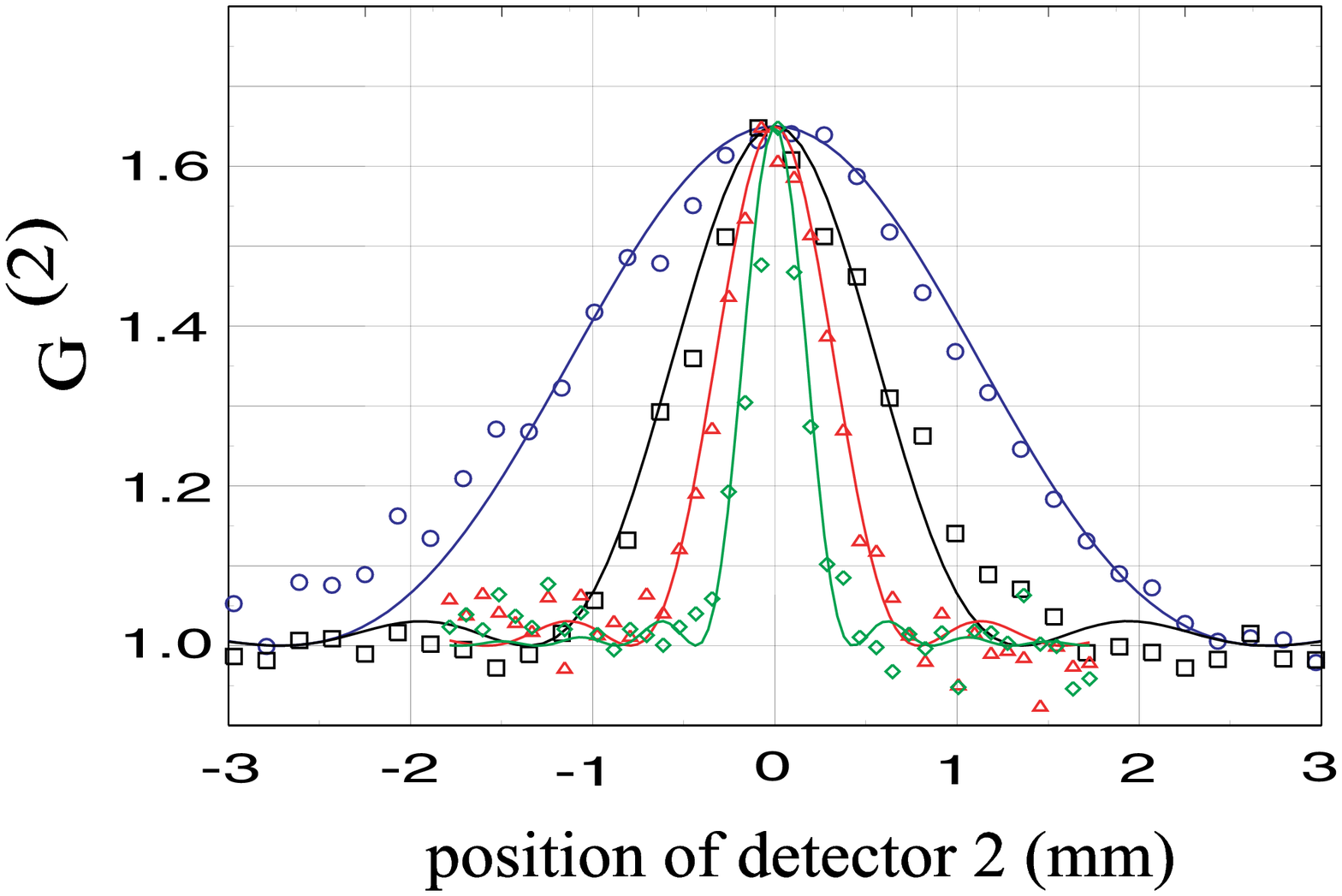}} \vspace{1cm}
Figure \ref{spotsize}.  Giuliano Scarcelli, Alejandra Valencia,
and Yanhua Shih.

\centerline{\epsfxsize=2in \epsffile{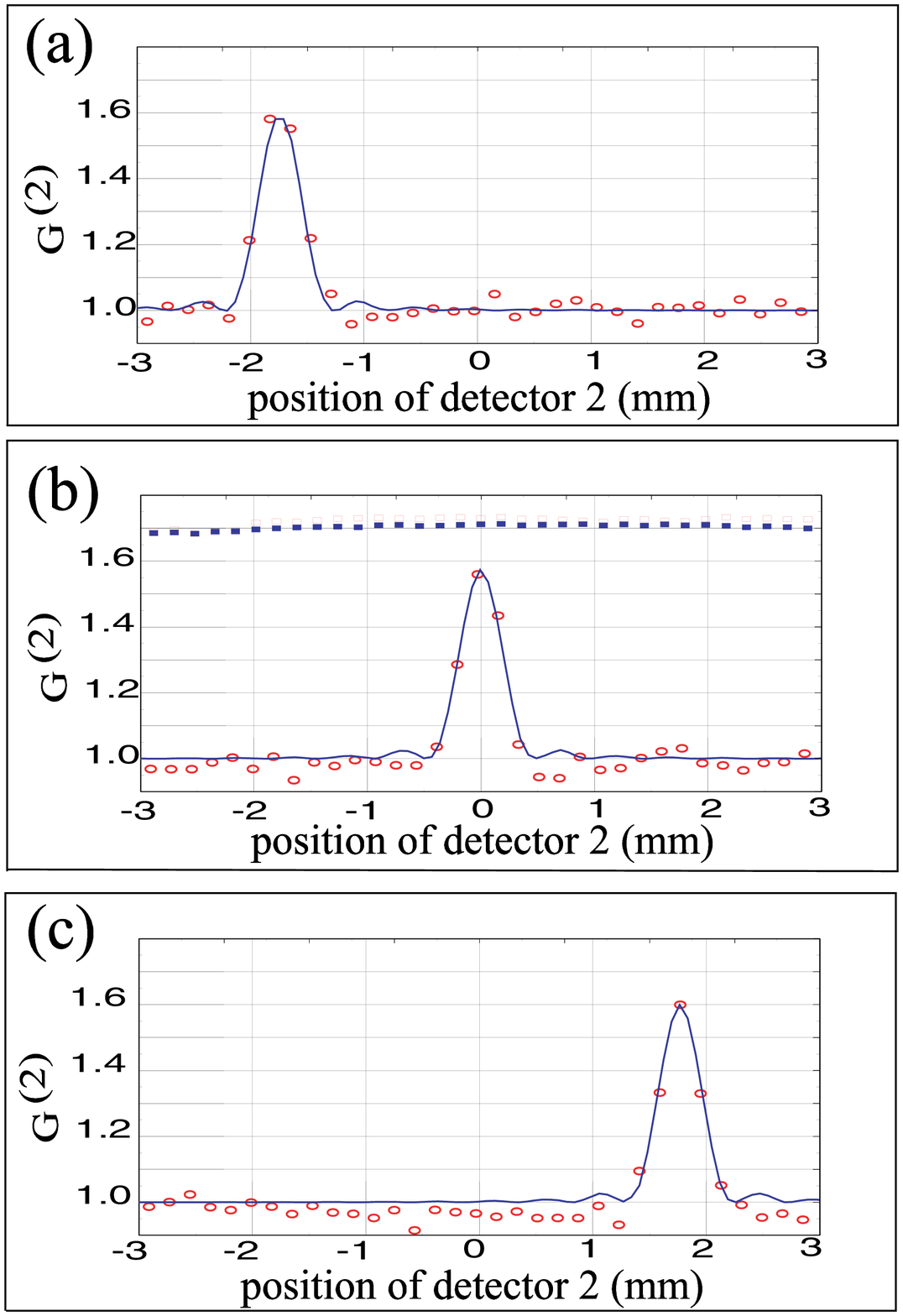}} \vspace{1cm}
Figure \ref{coin}.  Giuliano Scarcelli, Alejandra Valencia, and
Yanhua Shih.

\end{document}